\documentclass[eqsecnum,nofootinbib,12pt]{revtex4}
\usepackage{amsmath}
   \usepackage{bm}
\usepackage{amsthm,amscd}
\usepackage{mathrsfs}
\usepackage{verbatim}
\usepackage{amsfonts,amsmath,latexsym,amssymb,txfonts,bm}
\usepackage[american]{babel}
\usepackage{dsfont}

\begin{document}
\title{Small Mass Expansion of Functional Determinants on the Generalized Cone}
\author{
Guglielmo Fucci\footnote{Electronic address: Guglielmo\textunderscore Fucci@Baylor.edu} and Klaus Kirsten\footnote{Electronic address: Klaus\textunderscore Kirsten@Baylor.edu}
\thanks{Electronic address: gfucci@nmt.edu}}
\affiliation{Department of Mathematics, Baylor University, Waco, TX 76798 USA
}
\date{\today}
\vspace{2cm}
\begin{abstract}

In this paper we compute the small mass expansion for the functional determinant
of a scalar Laplacian defined on the bounded, generalized cone. In the framework of
zeta function regularization, we obtain an expression for the functional determinant
valid in any dimension for both Dirichlet and Robin boundary conditions in terms of the
spectral zeta function of the base manifold. Moreover, as a particular case, we specify the base
to be a $d$-dimensional sphere and present explicit results for $d=2,3,4,5$.

\end{abstract}
\maketitle
\section{Introduction}

The study of functional determinants of elliptic second order partial differential operators is of major importance in mathematical
physics and quantum field theory \cite{dewitt65,dewitt03,kirsten01}. In mathematics they encode particular important information about the
spectrum of the operator under consideration. In quantum field theory, instead, functional determinants of elliptic operators are used
in order to evaluate the one-loop effective action \cite{schwinger54,dewitt65,dewitt03,kirsten01,elizalde94}.
In the majority of cases of physical interest, one has to deal with second order hyperbolic linear partial differential
operators describing the dynamics of fields in Minkowski spacetime. By performing a Wick rotation to Euclidian
spacetime, the dynamical operator becomes elliptic. The mathematical advantage lies in the fact that
for elliptic self-adjoint partial differential operators on compact manifolds the spectral theorem holds.
This means, in particular, that the discrete spectrum is bounded from below and that all the eigenvalues can be ordered as follows
$-|C|\leq \lambda_{1}\leq \lambda_{2}\cdots\to+\infty$. One can then construct the spectral $\zeta$-function of the operator, say $L$,
as
\begin{equation}\label{0}
\zeta(s)=\sum_{n=1}^\infty\lambda_{n}^{-s}\;,
\end{equation}
which is convergent for $\textrm{Re}\,s> D/2$, with $D$ being the dimension of the manifold under consideration.
One can analytically continue, in a unique way, $\zeta(s)$ to a meromorphic function in the whole
complex plane which coincides with (\ref{0}) in its domain of convergence.

The functional determinant of the operator $L$ is defined as a product of all its eigenvalues.
Obviously such an expression is divergent and, thus, needs to be regularized. One way of making sense of this
infinite product is using $\zeta$-function regularization. In this framework, one \emph{defines}
the functional determinant of $L$ to be
\begin{equation}
\textrm{Det}\,L=\exp{[-\zeta^{\prime}(0)]}\;,
\end{equation}
where the derivative of $\zeta(s)$ at $s=0$ has been obtained by analytic continuation
in the way mentioned before. This object is of fundamental importance in quantum field theory
because it yields the one-loop effective action $\Gamma_{(1)}$ as
\begin{equation}
\Gamma_{(1)}=\sigma\ln\textrm{Det}\,L\;,
\end{equation}
where $\sigma=1/2,-1/2$ according to whether we are dealing with neutral scalar fields or Dirac fields.
In this paper we will utilize $\zeta$-function
techniques in order to obtain the small mass expansion of the functional determinant for the
scalar Laplacian on the generalized cone. This work represents an extension
to the massive case of the investigations initiated in the paper \cite{bordag96}; see also \cite{bordag96b,stuart}.
Apart from the mathematical and theoretical interest of computing small mass corrections
to the massless functional determinant, the main physical interest lies in the evaluation of the
one-loop effective action or the Casimir energy for massive fields on orbifolds.
In fact an orbifold is defined locally as the quotient space of a smooth manifold $X$ and a discrete
isometry group $G$ acting linearly on $X$. In general the action of the group $G$ on
$X$ will have fixed points, these points are then mapped to conical singularities in the quotient space.
We can thus say that orbifolds can be locally represented as generalized cones. These geometries are of
fundamental importance in order to compute the one-loop effective action in field
theoretical models containing orbifold compactification (see e.g. \cite{vongersdorff08}). It is hoped that by combining
a systematic small mass expansion with a large mass expansion from the heat kernel the intermediate range of mass values for
which serious numerical work is needed can be minimized. That this can be achieved in principle has recently been demonstrated
convincingly in \cite{min} in the context of the QCD instanton determinant.

The outline of the paper is as follows. In Section II we describe the geometry of a bounded generalized cone and we introduce the basic objects of our study. In particular eigenmodes, eigenvalues and the related zeta function of the Laplacian on the cone are discussed. We consider two boundary conditions, namely Dirichlet and Robin, and the functional determinants for massive scalar fields on the cone are evaluated for each case. Specializing the generalized cone to the case of the ball very explicit answers involving the zeta function of Riemann are given. The Appendix contains a list of polynomials needed for the computation and the Conclusions point to the most important results of the article.

\section{Geometric Background and $\zeta$-function}

In this paper we will consider a particular bounded manifold which is known as the generalized cone.
The generalized cone is defined as the $D=(d+1)$-dimensional manifold $\mathscr{M}=I\times\mathscr{N}$
where $\mathscr{N}$ is the base manifold, supposed to be a smooth Riemannian manifold possibly with boundary, and $I=[0,1]\subseteq \mathds{R}$. The generalized
cone is endowed with the hyperspherical metric \cite{cheeger83}
\begin{equation}\label{1}
ds^{2}=dr^{2}+r^{2}d\Sigma^{2}\;,
\end{equation}
where $d\Sigma^{2}$ represents the metric on $\mathscr{N}$ and $r\in I$.
It is known \cite{bordag96} that the curvatures on $\mathscr{M}$ and on the base $\mathscr{N}$
are conformally related as follows
\begin{eqnarray}\label{2}
R^{ij}{}_{kl}&=&\frac{1}{r^{2}}\left[\hat{R}^{ij}{}_{kl}-\left(\delta^{i}{}_{k}\delta^{j}{}_{l}-\delta^{i}{}_{l}\delta^{j}{}_{k}\right)\right]\;,\quad
R^{i}{}_{j}=\frac{1}{r^{2}}\left[\hat{R}^{i}{}_{j}-(d-1)\delta^{i}{}_{j}\right]\;,\\
R&=&\frac{1}{r^{2}}\left[\hat{R}-d(d-1)\right]\;,
\end{eqnarray}
where $R$ and $\hat{R}$ are the curvature tensors, respectively, on $\mathscr{M}$ and $\mathscr{N}$.
It is easily seen, from the relations (\ref{2}), that in general the manifold under consideration has a
singularity at the origin $r=0$.

Let $\Delta_{\mathscr{M}}$ be the Laplacian defined on the manifold $\mathscr{M}$ acting on
scalar functions $\varphi$. We are interested in the the following eigenvalue problem
\begin{equation}\label{3}
\left(-\Delta_{\mathscr{M}}+m^{2}\right)\varphi=\alpha^{2}\varphi\;,
\end{equation}
where the parameter $m$ represents the mass of the scalar field. In hyperspherical coordinates,
the Laplacian is separable and can be written as
\begin{equation}\label{4}
\Delta_{\mathscr{M}}=\frac{\partial^{2}}{\partial r^{2}}+\frac{d}{r}\frac{\partial}{\partial r}+\frac{1}{r^{2}}\Delta_{\mathscr{N}}\;,
\end{equation}
where $\Delta_{\mathscr{N}}$ denotes the Laplacian on the base manifold $\mathscr{N}$.
The solution of the eigenvalue equation (\ref{3}) which is regular at the origin can
be written as a product of a radial function and an angular one as follows
\begin{equation}\label{5}
\varphi=r^{\frac{1-d}{2}}J_{\nu}(\gamma r)Y(\Omega)\;,
\end{equation}
where $J_{\nu}$ is the Bessel function of the first kind and we have set $\alpha^{2}=\gamma^{2}+m^{2}$.
[Allowing for functions that are square integrable but singular at the origin results in the possibility of families of self-adjoint extensions of $\Delta_{\mathscr{M}}$ \cite{paul}, a topic not considered in this article.]
The angular part of (\ref{5}) is the solution of the equation
\begin{equation}\label{6}
\Delta_{\mathscr{N}}Y(\Omega)=-\lambda^{2}Y(\Omega)\;.
\end{equation}
At this point it is straightforward to show that the index $\nu$ of the Bessel
function satisfies the equation
\begin{equation}\label{7}
\nu^{2}=\lambda^{2}+\frac{(1-d)^{2}}{4}\;.
\end{equation}

The introduction of a constant curvature on the manifold $\mathscr{M}$ does not
add major complications. In fact, by denoting with $\xi$ the coupling constant,
the eigenvalue equation (\ref{3}) acquires the term $\xi R$. In this
situation the index $\nu$ becomes
\begin{equation}\label{8}
\nu^{2}=\lambda^{2}+\xi\hat{R}+d(d-1)(\xi_{d}-\xi)\;,
\end{equation}
where $\xi_{d}=(d-1)/4d$ is the conformal coupling constant in $d$ dimensions. The case in which
the curvature $\hat{R}$ is not constant has been described in \cite{bordag96}.

Let us now turn our attention to the spectral $\zeta$-function associated with $-\Delta_{\mathscr{M}}+m^2$ on the manifold $\mathscr{M}$.
This is defined as
\begin{equation}\label{9}
\zeta_{\mathscr{M}}(s)=\sum_\alpha \left(\alpha^{2}+m^{2}\right)^{-s}\;,
\end{equation}
where we will assume that no negative eigenvalues occur so that we can use the standard branch cut of the logarithm.
Our aim is to express the zeta function for the whole manifold $\mathscr{M}$
as much as possible in terms of the base manifold zeta function $\zeta_{\mathscr{N}}$ \cite{cheeger83} which is
defined as
\begin{equation}\label{10}
\zeta_{\mathscr{N}}(s)=\sum_{\nu} d(\nu)\nu^{-2s}\;,
\end{equation}
where $d(\nu)$ is the degeneracy of the scalar harmonics $Y(\Omega)$ on $\mathscr{N}$.
Without specifying the base manifold $\mathscr{N}$ one is still able to impose
boundary conditions \cite{bordag96}, in particular we are interested in setting
\begin{equation}\label{11}
J_{\nu}(\gamma )=0\;,
\end{equation}
for Dirichlet, and
\begin{equation}\label{12}
\left(1-\frac{d+1}{2}-\beta\right)J_{\nu}(\gamma )+\gamma  J^{\prime}_{\nu}(\gamma )=0\;,
\end{equation}
for Robin. Notice that for $\beta=0$ the last relation reduces to Neumann boundary condition.
For the problem under consideration the eigenvalues of the Laplacian are not known explicitly,
however the boundary conditions (\ref{11}) and (\ref{12}) allow their determination in an implicit
fashion.

For Dirichlet boundary conditions a convenient way to express $\zeta_{\mathscr{M}}(s)$ in terms of a contour integral in the complex
plane is as follows \cite{bordag96,bordag96a,bordag96b,esposito97,kirsten01},
\begin{equation}\label{13}
\zeta_{\mathscr{M}}(s)=\sum_\nu d(\nu)\frac{1}{2\pi i}\int_{\Gamma}dk\left[k^{2}+m^{2}\right]^{-s}\frac{\partial}{\partial k}\ln J_{\nu}(k)\;,
\end{equation}
where $\Gamma$ is a contour that encircles all the zeroes of $J_{\nu}(k)$ on the positive real axis in the counterclockwise direction.
By deforming the contour $\Gamma$ to the imaginary axis one gets \cite{bordag96b,kirsten01}
\begin{equation}\label{14}
\zeta_{\mathscr{M}}(s)=\sum_\nu d(\nu)\frac{\sin(\pi s)}{\pi}\int_{m}^{\infty}dk\left[k^{2}-m^{2}\right]^{-s}\frac{\partial}{\partial k}\ln \left(k^{-\nu}J_{\nu}(k)\right)\;.
\end{equation}
At this point, it is very useful to split the spectral $\zeta$-function into two parts \cite{bordag96a,bordag96b,kirsten01}.
In order to do so we exploit the asymptotic expansion of the modified Bessel functions $I_{\nu}(k)$ for
$\nu\to\infty$ and for $z=k/\nu$ fixed as \cite{olver54,erdelyi53}
\begin{equation}\label{15}
I_{\nu}(\nu z)\sim\frac{1}{\sqrt{2\pi\nu}}\frac{e^{\nu\eta}}{(1+z^{2})^{1/4}}\left[1+\sum_{k=1}^{\infty}\frac{u_{k}(t)}{\nu^{k}}\right]\;,
\end{equation}
where the polynomials $u_{k}(t)$ are determined by the recurrence relation
\begin{equation}\label{16}
u_{k+1}(t)=\frac{1}{2}t^{2}(1-t^{2})u_{k}^{\prime}(t)+\frac{1}{8}\int_{0}^{t}d\tau(1-5\tau^{2})u_{k}(\tau)\;,
\end{equation}
with $u_{0}(t)=1$ and
\begin{equation}\label{17}
t=\frac{1}{\sqrt{1+z^{2}}}\;,\qquad \eta=\sqrt{1+z^{2}}+\ln\left[\frac{z}{1+\sqrt{1+z^{2}}}\right]\;.
\end{equation}
By adding and subtracting $N$ leading terms of the asymptotic expansion (\ref{15}) one can write \cite{bordag96,bordag96a,bordag96b,kirsten01}
\begin{equation}\label{18}
\zeta_{\mathscr{M}}(s)=Z(s)+\sum_{i=-1}^{N}A_{i}(s)\;,
\end{equation}
where
\begin{eqnarray}\label{19}
A_{-1}(s)&=&\frac{1}{4\sqrt{\pi}\,\Gamma(s)}\sum_{j=0}^{\infty}\frac{(-1)^{j}}{j!}\frac{\Gamma\left(s+j+\frac{1}{2}\right)}{s+j}m^{2j}\zeta_{\mathscr{N}}\left(s+j-\frac{1}{2}\right)\;,\\
A_{0}(s)&=&-\frac{1}{4}\sum_{j=0}^{\infty}\frac{(-1)^{j}}{j!}\frac{\Gamma\left(s+j\right)}{\Gamma(s)}m^{2j}\zeta_{\mathscr{N}}\left(s+j\right)\;,\\
A_{i}(s)&=&-\frac{1}{\Gamma(s)}\sum_{j=0}^{\infty}\frac{(-1)^{j}}{j!}m^{2j}\zeta_{\mathscr{N}}\left(s+j+\frac{i}{2}\right)\sum_{b=0}^{i}x_{i,b}\frac{\Gamma\left(s+b+j+\frac{i}{2}\right)}{\Gamma\left(b+\frac{i}{2}\right)}\;,\label{19a}
\end{eqnarray}
and
\begin{equation}
Z(s)=\sum_{\nu} d(\nu)Z_{\nu}(s)\;,
\end{equation}
with
\begin{eqnarray}\label{20}
Z_{\nu}(s)&=&\frac{\sin(\pi s)}{\pi}\int_{\frac{m}{\nu}}^{\infty}dz\,\,[(z\nu)^{2}-m^{2}]^{-s}\\
& &\hspace{2.0cm}\times\frac{\partial}{\partial z}
\Bigg\{\ln\left[z^{-\nu}I_{\nu}(z\nu)\right]
-\ln\left[\frac{z^{-\nu}}{\sqrt{2\pi\nu}}\frac{e^{\nu\eta}}{(1+z^{2})^{1/4}}\right]
-\sum_{n=1}^{N}\frac{D_{n}(t)}{\nu^{n}}\Bigg\}\;.\;\;\;\;\;\;\;\;\; \nonumber
\end{eqnarray}
The terms $D_{n}(t)$ appearing in (\ref{20}) are defined through the cumulant expansion \cite{bordag96,bordag96a,bordag96b,kirsten01}
\begin{equation}\label{21}
\ln\left[1+\sum_{k=1}^{\infty}\frac{u_{k}(t)}{\nu^{k}}\right]\sim\sum_{n=1}^{\infty}\frac{D_{n}(t)}{\nu^{n}}\;,
\end{equation}
and have the polynomial structure
\begin{equation}\label{22}
D_{n}(t)=\sum_{i=0}^{n}x_{i,n}t^{n+2i}\;.
\end{equation}
The polynomials (\ref{22}) are listed, up to the sixth order, in Appendix \ref{app1}.

\section{Small Mass Expansion for Dirichlet Boundary Conditions}

We are particularly interested in the evaluation of the first mass correction to the
functional determinant of the scalar Laplacian on the generalized cone. Higher orders can be obtained along the same lines.
In this case, the relevant formulas for
the terms (\ref{19})-(\ref{19a}) are
\begin{eqnarray}
A_{-1}(s)&=&\frac{1}{4\sqrt{\pi}}\frac{\Gamma\left(s-\frac{1}{2}\right)}{\Gamma(s+1)}\left\{\zeta_{\mathscr{N}}\left(s-\frac{1}{2}\right)-m^{2}\frac{s\left(s-\frac{1}{2}\right)}{s+1}\zeta_{\mathscr{N}}\left(s+\frac{1}{2}\right)\right\}+O(m^{4})\;,\\
A_{0}(s)&=&-\frac{1}{4}\left\{\zeta_{\mathscr{N}}(s)-m^{2}s\zeta_{\mathscr{N}}(s+1)\right\}+O(m^{4})\;,\\
A_{i}(s)&=&-\frac{1}{\Gamma(s)}\sum_{b=0}^{i}x_{i,b}\frac{\Gamma\left(s+b+\frac{i}{2}\right)}{\Gamma\left(b+\frac{i}{2}\right)}\left\{\zeta_{\mathscr{N}}\left(s+\frac{i}{2}\right)-m^{2}\left(s+b+\frac{i}{2}\right)\zeta_{\mathscr{N}}\left(s+1+\frac{i}{2}\right)\right\}+O(m^{4})\;.\;\;\;\;\;\;\;\;\label{22a}
\end{eqnarray}

It is not difficult to see, in equation (\ref{18}), that in order to evaluate the functional determinant
we need to compute the derivatives of the quantities (\ref{19})-(\ref{20}). We would like to point out here that for the explicit calculation that will follow
it is sufficient to subtract the first $D-1=d$ terms of the asymptotic expansion in (\ref{15}) \cite{bordag96,bordag96a,bordag96b}. Therefore, we will set $N=d$
in equation (\ref{18}).

Let us start with the
function $Z_{\nu}(s)$. By performing the derivative and then setting $s=0$ one obtains
\begin{eqnarray}
Z_{\nu}^{\prime}(0)&=&-\ln\,I_{\nu}(m)-\frac{1}{2}\ln    2\pi\nu-\frac{1}{4}\ln\left(1+\frac{m^{2}}{\nu^{2}}\right)+\nu\sqrt{1+\frac{m^{2}}{\nu^{2}}}+\frac{\nu}{2}\ln\frac{m^{2}}{\nu^{2}}\nonumber\\
&-&\nu\ln{\left(1+\sqrt{1+\frac{m^{2}}{\nu^{2}}}\right)}+\sum_{n=1}^{d}\nu^{-n}D_{n}\left(\left(1+\frac{m^{2}}{\nu^{2}}\right)^{-\frac{1}{2}}\right)\;.
\end{eqnarray}
By expanding the last expression in terms of the mass up to the order $m^{2}$, and by noticing that
\begin{equation}
\sum_{n=1}^{d}\nu^{-n}D_{n}\left(\left(1+\frac{m^{2}}{\nu^{2}}\right)^{-\frac{1}{2}}\right)
=\sum_{n=1}^{d}\frac{D_{n}(1)}{\nu^{n}}-m^{2}\sum_{n=1}^{d}\nu^{-n-2}\left[\frac{nD_{n}(1)}{2}+\sum_{b=0}^{n}x_{n,b}b\right]+O(m^{4})\;,
\end{equation}
we get
\begin{eqnarray}
Z_{\nu}^{\prime}(0)&=&\ln\Gamma(\nu+1)+\nu-\nu\ln\nu-\frac{1}{2}\ln 2\pi\nu+\sum_{n=1}^{d}\frac{D_{n}(1)}{\nu^{n}}\nonumber\\
&-&m^{2}\left\{\frac{1}{4(\nu+1)}-\frac{1}{4\nu}+\frac{1}{4\nu^{2}}+\frac{1}{2\nu^{2}}\sum_{n=1}^{d}\frac{D^{\prime}_{n}(1)}{\nu^{n}}\right\}+O(m^{4})\;,
\end{eqnarray}
where $D^{\prime}_{n}(1)$ denotes the derivative of the polynomials $D_{n}(t)$ evaluated at $t=1$, and we have used the fact that
\begin{equation}
\frac{nD_{n}(1)}{2}+\sum_{b=0}^{n}x_{n,b}b=\frac{1}{2}D^{\prime}_{n}(1)\;.
\end{equation}

In what follows we will write the spectral $\zeta$-function on the base manifold $\mathscr{N}$
in a way that explicitly shows its structure, namely we have
\begin{equation}\label{23}
\zeta_{\mathscr{N}}(s+\alpha)=\frac{1}{s}\textrm{Res}\,\zeta_{\mathscr{N}}(\alpha)+\textrm{FP}\,\zeta_{\mathscr{N}}(\alpha)+O(s)\;,
\end{equation}
and for its derivative
\begin{equation}\label{24}
\zeta^{\prime}_{\mathscr{N}}(s+\alpha)=-\frac{1}{s^{2}}\textrm{Res}\,\zeta_{\mathscr{N}}(\alpha)+\zeta^{\prime}_{\mathscr{N}}(\alpha)+O(s)\;,
\end{equation}
where $\textrm{Res}$ denotes the residue of the function and $\textrm{FP}$ its finite part.
We would like to make an important remark here. As already mentioned earlier, the generalized cone
is a singular manifold and for this reason the heat kernel asymptotic expansion in general
will contain a non-standard logarithmic term \cite{bruning87}. This behavior is translated into the appearance
of a pole at $s=0$ in the spectral $\zeta$-function on $\mathscr{M}$ \cite{bordag96b,cheeger83,voros87}. In what follows,
we will assume that the logarithmic term in the heat kernel asymptotic expansion does not appear, this will allow a
standard definition of the functional determinant \cite{bordag96b}. From a more formal point of
view, as said, this means that $\zeta_{\mathscr{M}}(s)$ is regular at $s=0$, which is the case if $\zeta_{\mathscr{N}} (s)$ is regular at $s=-1/2$.

By using the expressions (\ref{23}) and (\ref{24}), it is not difficult to obtain the derivative of $A_{-1}(s)$ at $s=0$;
more specifically we have
\begin{equation}\label{25}
A^{\prime}_{-1}(0)=\left(\ln 2-1\right)\zeta_{\mathscr{N}}\left(-\frac{1}{2}\right)-\frac{1}{2}\zeta^{\prime}_{\mathscr{N}}\left(-\frac{1}{2}\right)+m^{2}\left[\frac{1}{4}(2\ln 2+1)\textrm{Res}\,\zeta_{\mathscr{N}}\left(\frac{1}{2}\right)-\frac{1}{4}\textrm{FP}\,\zeta_{\mathscr{N}}\left(\frac{1}{2}\right)\right]+O(m^{4})\;.\;\;\;\;\;\;\;
\end{equation}
For the derivative at $s=0$ of $A_{0}(s)$ one obtains
\begin{equation}\label{26}
A^{\prime}_{0}(0)=-\frac{1}{4}\zeta^{\prime}_{\mathscr{N}}(0)+\frac{m^{2}}{4}\textrm{FP}\,\zeta_{\mathscr{N}}(1)+O(m^{4})\;,
\end{equation}
and, finally, for the terms $A_{i}(s)$ in (\ref{22a}) we get
\begin{eqnarray}\label{27}
A_{i}^{\prime}(0)&=&-\frac{\zeta_{R}(-i)}{i}\left[\gamma\textrm{Res}\,\zeta_{\mathscr{N}}\left(\frac{i}{2}\right)+\textrm{FP}\,\zeta_{\mathscr{N}}\left(\frac{i}{2}\right)\right]-\sum_{b=0}^{i}x_{i,b}\Psi\left(b+\frac{i}{2}\right)\textrm{Res}\,\zeta_{\mathscr{N}}\left(\frac{i}{2}\right)\nonumber\\
&+&m^{2}\Bigg\{\frac{\zeta_{R}(-i)}{i}\textrm{Res}\,\zeta_{\mathscr{N}}\left(\frac{i}{2}+1\right)+\sum_{b=0}^{i}x_{i,b}\left(b+\frac{i}{2}\right)\Bigg[\left(\gamma+\Psi\left(b+\frac{i}{2}\right)\right)\textrm{Res}\,\zeta_{\mathscr{N}}\left(\frac{i}{2}+1\right)\nonumber\\
&+&\textrm{FP} \,\zeta_{\mathscr{N}}\left(\frac{i}{2}+1\right)\Bigg]\Bigg\}+O(m^{4})\;,
\end{eqnarray}
where $\gamma$ is the Euler-Mascheroni constant, $\zeta_{R}(s)$ represents the Riemann $\zeta$-function and $\Psi(s)$ is the logarithmic derivative of the
Gamma function.

At this point, we exploit the integral representation of the function $\ln\Gamma(\nu)$ \cite{gradshtein07},
\begin{equation}\label{28}
\ln\Gamma(\nu)=\left(\nu-\frac{1}{2}\right)\ln\nu-\nu+\frac{1}{2}\ln 2\pi+\int_{0}^{\infty}dt\left(\frac{1}{2}-\frac{1}{t}+\frac{1}{e^{t}-1}\right)\frac{e^{-t\nu}}{t}\;,
\end{equation}
and the integral representation for the inverse powers of $\nu$ as follows
\begin{equation}\label{29}
\nu^{-n}=\frac{1}{\Gamma(n)}\int_{0}^{\infty}dt\,t^{n-1}e^{-t\nu}\;,
\end{equation}
to find
\begin{eqnarray}\label{30}
Z_{\nu}^{\prime}(0)&=&\int_{0}^{\infty}dt\left[\sum_{n=1}^{d}\frac{D_{n}(1)}{\Gamma(n)}t^{n}+\frac{1}{2}-\frac{1}{t}+\frac{1}{e^{t}-1}\right]\frac{e^{-t\nu}}{t}\nonumber\\
&-&m^{2}\int_{0}^{\infty}dt\left(\sum_{n=1}^{d}\frac{D_{n}^{\prime}(1)}{2\Gamma(n+2)}t^{n+2}+\frac{t e^{-t}}{4}-\frac{t}{4}+\frac{t^{2}}{4}\right)\frac{e^{-t\nu}}{t}+O(m^{4})\;.
\end{eqnarray}

It will be convenient to define the "square root"
heat kernel associated with $\nu$ \cite{bordag96}
\begin{equation}\label{31}
K_{\mathscr{N}}^{1/2}(t)=\sum d(\nu)e^{-t\lambda}\;.
\end{equation}
By introducing a regularization parameter $z$ and by utilizing the definition (\ref{31}), one can rewrite (\ref{30}) as
\begin{eqnarray}\label{32}
Z^{\prime}(0,z)&=&\int_{0}^{\infty}dt\,t^{z-1}\left[\sum_{n=1}^{d}\frac{D_{n}(1)}{\Gamma(n)}t^{n}+\frac{1}{2}-\frac{1}{t}+\frac{1}{e^{t}-1}\right]K_{\mathscr{N}}^{1/2}(t)\nonumber\\
&-&m^{2}\int_{0}^{\infty}dt\,t^{z-1}\left(\sum_{n=1}^{d}\frac{D_{n}^{\prime}(1)}{2\Gamma(n+2)}t^{n+2}+\frac{t e^{-t}}{4}-\frac{t}{4}+\frac{t^{2}}{4}\right)K_{\mathscr{N}}^{1/2}(t)+O(m^{4})\;.
\end{eqnarray}

By recalling that the spectral $\zeta$-function is obtained from the heat kernel by an inverse Mellin transform as \cite{esposito97,dewitt03,kirsten01,vassile03}
\begin{equation}
\zeta_{\mathscr{N}}\left(\frac{s}{2}\right)=\frac{1}{\Gamma(s)}\int_{0}^{\infty}dt\,t^{s-1}K_{\mathscr{N}}^{1/2}(t)\;,
\end{equation}
we can express the function (\ref{32}) in the form
\begin{eqnarray}\label{33}
Z^{\prime}(0,z)&=&\sum_{n=1}^{d}\frac{D_{n}(1)}{\Gamma(n)}t^{n}\Gamma(z+n)\zeta_{\mathscr{N}}\left(\frac{z+n}{2}\right)+\frac{1}{2}\Gamma(z)\zeta_{\mathscr{N}}\left(\frac{z}{2}\right)
-\Gamma(z-1)\zeta_{\mathscr{N}}\left(\frac{z-1}{2}\right)+\Gamma(z)\zeta_{\mathscr{N}+1}\left(z\right)\nonumber\\
&-&m^{2}\Bigg[\sum_{n=1}^{d}\frac{D_{n}^{\prime}(1)}{2\Gamma(n+2)}t^{n+2}\Gamma(z+n+2)\zeta_{\mathscr{N}}\left(\frac{z+n+2}{2}\right)-\frac{1}{4}\Gamma(z+1)\zeta_{\mathscr{N}}\left(\frac{z+1}{2}\right)\nonumber\\
&+&\frac{1}{4}\Gamma(z+2)\zeta_{\mathscr{N}}\left(\frac{z+2}{2}\right)+\frac{1}{4}\Gamma(z+1)\zeta_{\mathscr{N}}(z+1,1)\Bigg]\;,
\end{eqnarray}
where we have introduced
\begin{equation}
\zeta_{\mathscr{N}+1}(z)=\frac{1}{\Gamma(z)}\int_{0}^{\infty}t^{z-1}\frac{K_{\mathscr{N}}^{1/2}(t)}{e^{t}-1}\;,
\end{equation}
and
\begin{equation}\label{33b}
\zeta_{\mathscr{N}}(z,u)=\frac{1}{\Gamma(z)}\sum d(\nu)\int_{0}^{\infty}dt\,t^{z-1}e^{-(\nu+u)t}=\frac{1}{\Gamma(z)}\int_{0}^{\infty}dt\,t^{z-1}e^{-ut}K_{\mathscr{N}}^{1/2}(t)\;.
\end{equation}

The expression of primary interest, namely $Z^{\prime}(0)$, is obtained by just taking the limit as $z$ approaches zero
of (\ref{33}). By recalling the structure of the spectral $\zeta$-function on $\mathscr{N}$ in (\ref{23}) and (\ref{24}),
the limit as $z\to 0$ gives
\begin{eqnarray}\label{34}
Z^{\prime}(0)&=&\sum_{n=1}^{d}\frac{\zeta_{R}(-n)}{n}\left[2\Psi(n)\textrm{Res}\,\zeta_{\mathscr{N}}\left(\frac{n}{2}\right)+\textrm{FP}\,\zeta_{\mathscr{N}}\left(\frac{n}{2}\right)\right]+
(1-\gamma)\zeta_{\mathscr{N}}\left(-\frac{1}{2}\right)-\frac{\gamma}{2}\zeta_{\mathscr{N}}\left(0\right)+\frac{1}{4}\zeta_{\mathscr{N}}^{\prime}\left(0\right)\nonumber\\
&+&\frac{1}{2}\zeta_{\mathscr{N}}^{\prime}\left(-\frac{1}{2}\right)+\lim_{z\to 0}\Bigg\{\frac{2}{z}\sum_{n=1}^{d}\frac{\zeta_{R}(-n)}{n}\textrm{Res}\,\zeta_{\mathscr{N}}\left(\frac{n}{2}\right)
+\frac{1}{2z}\zeta_{\mathscr{N}}\left(0\right)+\frac{1}{z}\zeta_{\mathscr{N}}\left(-\frac{1}{2}\right)+\Gamma(z)\zeta_{\mathscr{N}+1}\left(z\right)\Bigg\}\nonumber\\
&-&m^{2}\Bigg\{\sum_{n=1}^{d}\frac{D^{\prime}(1)}{2}\left[2\Psi(n+2)\textrm{Res}\,\zeta_{\mathscr{N}}\left(\frac{n+2}{2}\right)+\textrm{FP}\,\zeta_{\mathscr{N}}\left(\frac{n+2}{2}\right)\right]
+\frac{\gamma}{2}\textrm{Res}\,\zeta_{\mathscr{N}}\left(\frac{1}{2}\right)\nonumber\\
&+&\frac{1-\gamma}{2}\textrm{Res}\,\zeta_{\mathscr{N}}\left(1\right)-\frac{1}{4}\textrm{FP}\,\zeta_{\mathscr{N}}\left(\frac{1}{2}\right)+\frac{1}{4}\textrm{FP}\,\zeta_{\mathscr{N}}\left(1\right)
+\lim_{z\to 0}\Bigg[\sum_{n=1}^{d}\frac{D^{\prime}_{n}(1)}{z}\textrm{Res}\,\zeta_{\mathscr{N}}\left(\frac{n+2}{2}\right)\nonumber\\
&-&\frac{1}{2z}\textrm{Res}\,\zeta_{\mathscr{N}}\left(\frac{1}{2}\right)+\frac{1}{2z}\textrm{Res}\,\zeta_{\mathscr{N}}\left(1\right)+\frac{1}{4}\Gamma(z+1)\zeta_{\mathscr{N}}(z+1,1)
\Bigg]
\Bigg\}+O(m^{4})\;.
\end{eqnarray}

As a last step, we have to explicitly compute the remaining limits in equation (\ref{34}). By utilizing the
series \cite{bordag96,gradshtein07}
\begin{equation}
\frac{1}{e^{t}-1}=\frac{1}{t}-\frac{1}{2}-\sum_{n=1}^{\infty}\frac{t^{n}}{n!}\zeta_{R}(-n)\;,
\end{equation}
and the asymptotic expansion of the heat kernel
\begin{equation}\label{34b}
K_{\mathscr{N}}^{1/2}(t)\sim\sum\tilde{a}_{k}t^{k-d}\;,
\end{equation}
where
\begin{equation}
\tilde{a}_{k}=2\Gamma(d-k)\textrm{Res}\,\zeta_{\mathscr{N}}\left(\frac{d-k}{2}\right)\;,
\end{equation}
for $k=0,1,\cdots ,d-1$, $k=d-(2l+1)$ with $l\in\mathds{N}^{+}$, and
\begin{equation}
\tilde{a}_{k}=\frac{(-1)^{k-d}}{(k-d)!}\zeta_{\mathscr{N}}\left(\frac{d-k}{2}\right)\;,
\end{equation}
for $k>d$ and $k\in\mathds{N}^{+}$, one can prove that \cite{bordag96}
\begin{equation}\label{35}
\zeta_{\mathscr{N}+1}(0)=-\zeta_{\mathscr{N}}\left(-\frac{1}{2}\right)-\frac{1}{2}\zeta_{\mathscr{N}}(0)-2\sum_{n=1}^{d}\frac{\zeta_{R}(-n)}{n}\textrm{Res}\,\zeta_{\mathscr{N}}\left(\frac{n}{2}\right).
\end{equation}
The value $\zeta_{\mathscr{N}+1}(0)$ just computed is needed in the small $z$ expansion
\begin{equation}\label{36}
\Gamma(z)\zeta_{\mathscr{N}+1}\left(z\right)= \frac{1}{z}\zeta_{\mathscr{N}+1}(0)-\gamma\zeta_{\mathscr{N}+1}(0)+\zeta^{\prime}_{\mathscr{N}+1}(0)+O(z)\;.
\end{equation}
By using this expansion together with (\ref{35}) in the part of zeroth order in the
mass of equation (\ref{34}), one can easily see that the
quantities containing the divergence $1/z$ cancel each other exactly.

In order to present the complete final result for Dirichlet boundary conditions, we need also to consider the terms of quadratic order in mass.
Exactly in the same way as we have done for the terms of order zero in the mass, we exploit the following small $z$ expansion
\begin{equation}
\Gamma(z+1)\zeta_{\mathscr{N}}(z+1,1)=\frac{1}{z}\textrm{Res}\,\zeta_{\mathscr{N}}(1,1)-\gamma\textrm{Res}\,\zeta_{\mathscr{N}}(1,1)+\textrm{FP}\,\zeta_{\mathscr{N}}(1,1)+O(z)\;.
\end{equation}
The value of $\textrm{Res}\,\zeta_{\mathscr{N}}(z+1,1)$ at $z=0$ can be easily obtained, in the same way as before, by using the
asymptotic expansion (\ref{34b}) and the definition (\ref{33b}). Explicitly, we get
\begin{equation}
\textrm{Res}\,\zeta_{\mathscr{N}}(1,1)=2\sum_{n=0}^{d}(-1)^{n}\textrm{Res}\,\zeta_{\mathscr{N}}\left(\frac{n+1}{2}\right)\;.
\end{equation}
By substituting this last expression in the $m^{2}$ terms containing $1/z$ in (\ref{34}), one easily finds
\begin{eqnarray}\label{37}
&&\lim_{z\to 0}\Bigg[\sum_{n=1}^{d}\frac{D^{\prime}_{n}(1)}{z}\textrm{Res}\,\zeta_{\mathscr{N}}\left(\frac{n+2}{2}\right)
-\frac{1}{2z}\textrm{Res}\,\zeta_{\mathscr{N}}\left(\frac{1}{2}\right)+\frac{1}{2z}\textrm{Res}\,\zeta_{\mathscr{N}}\left(1\right)+\frac{1}{4}\Gamma(z+1)\zeta_{\mathscr{N}}(z+1,1)
\Bigg]\nonumber\\
&=&\lim_{z\to 0}\Bigg[\frac{1}{z}\sum_{n=1}^{d}\left(D_{n}^{\prime}(1)-\frac{(-1)^{n}}{2}\right)\textrm{Res}\,\zeta_{\mathscr{N}}\left(\frac{n+2}{2}\right)\Bigg]\;.
\end{eqnarray}
It is straightforward to prove, see e.g. \cite{kirsten01}, that
\begin{equation}
D_{n}^{\prime}(1)=\frac{(-1)^{n}}{2}\;,
\end{equation}
and, therefore, the expression in (\ref{37}) vanishes identically. By combining the results in (\ref{25})-(\ref{27}) and (\ref{34})
we can finally obtain the expression for the small mass expansion of the functional determinant of the Laplacian
on the generalized cone as follows
\begin{eqnarray}\label{38}
\zeta^{\prime}_{\mathscr{M}}(0)&=&\zeta^{\prime}_{\mathscr{N}+1}(0)+2\sum_{n=1}^{d}\frac{\zeta_{R}(-n)}{n}\textrm{Res}\,\zeta_{\mathscr{N}}\left(\frac{n}{2}\right)\sum_{k=1}^{n-1}\frac{1}{k}+\ln2\,\zeta_{\mathscr{N}}\left(-\frac{1}{2}\right)\nonumber\\
&+&2\sum_{n=1}^{d}\textrm{Res}\,\zeta_{\mathscr{N}}\left(\frac{n}{2}\right)\int_{0}^{1}dt\,\frac{D_{n}(t)-t^{2}D_{n}(1)}{t(1-t^{2})}\nonumber\\
&+&m^{2}\Bigg[-\frac{1}{4}\textrm{FP}\zeta_{\mathscr{N}}(1,1)-\frac{1}{2}\textrm{Res}\,\zeta_{\mathscr{N}}(1)+\frac{1}{4}(2\ln 2+1)\textrm{Res}\,\zeta_{\mathscr{N}}\left(\frac{1}{2}\right)\nonumber\\
&+&\sum_{n=1}^{d}\frac{\zeta_{R}(-n)}{n}\textrm{Res}\,\zeta_{\mathscr{N}}\left(\frac{n}{2}+1\right)
-2\sum_{n=1}^{d}\frac{D^{\prime}_{n}(1)}{2}\textrm{Res}\,\zeta_{\mathscr{N}}\left(\frac{n}{2}+1\right)\left(\sum_{k=1}^{n-1}\frac{1}{k}+\frac{2n+1}{n(n+1)}\right)\nonumber\\
&-&\sum_{n=1}^{d}\textrm{Res}\,\zeta_{\mathscr{N}}\left(\frac{n}{2}+1\right)\int_{0}^{1}dt\,\frac{D_{n}^{\prime}(t)-tD_{n}^{\prime}(1)}{1-t^{2}}
\Bigg]+O(m^{4})\;,
\end{eqnarray}
where, in order to obtain the last expression, we have used the relations
\begin{equation}
\Psi\left(z+\frac{n}{2}\right)=-\gamma-2\int_{0}^{1}\frac{t^{2z+n}-t^{2}}{t(1-t^{2})}dt\;,
\end{equation}
and
\begin{equation}
\Psi(n+2)=-\gamma+\sum_{k=1}^{n-1}\frac{1}{k}+\frac{1}{n}+\frac{1}{n+1}\;.
\end{equation}
We would like to point out that the terms of order zero in the mass in equation (\ref{38})
coincide with the ones obtained in \cite{bordag96}. The terms proportional to $m^{2}$ are, instead, the \emph{new} massive corrections
to the scalar Laplacian on the generalized cone for Dirichlet boundary conditions.

\section{Small Mass Expansion for Robin Boundary Conditions}

The calculational procedure to follow in order to compute $\zeta_{\mathscr{M}}^{\prime}(0)$ for Robin boundary conditions
closely resembles the one used in the previous section for the Dirichlet case.
We will describe only the few necessary changes \cite{bordag96,bordag96a,bordag96b}. For Robin boundary conditions
we need, in addition to the asymptotic expansion (\ref{15}), the following one for $I^{\prime}_{\nu}(\nu z)$ \cite{gradshtein07,olver54}
\begin{equation}\label{38a}
I^{\prime}_{\nu}(\nu z)\sim \frac{1}{\sqrt{{2\pi\nu}}}\frac{e^{\nu\eta}(1+z^{2})^{1/4}}{z}\left[1+\sum_{k=1}^{\infty}\frac{v_{k}(t)}{\nu^{k}}\right]\;,
\end{equation}
where the polynomials $v_{k}(t)$ are determined by the recurrence relation
\begin{equation}\label{39}
v_{k}(t)=u_{k}(t)+t(t^{2}-1)\left[\frac{1}{2}u_{k-1}(t)+t u^{\prime}_{k-1}(t)\right]\;.
\end{equation}
In analogy with the Dirichlet case, we will make use of the cumulant expansion \cite{bordag96,bordag96a,bordag96b,kirsten01}
\begin{equation}\label{40}
\ln\left[1+\sum_{k=1}^{\infty}\frac{v_{k}(t)}{\nu^{k}}+\frac{u}{\nu}t\left(1+\sum_{k=1}^{\infty}\frac{u_{k}(t)}{\nu^{k}}\right)\right]\sim\sum_{n=1}^{\infty}\frac{M_{n}(t,u)}{\nu^{k}}\;,
\end{equation}
where the terms $M_{k}(t,u)$ have a polynomials structure analogous to the $D_{n}(t)$, namely
\begin{equation}\label{41}
M_{n}(t,u)=\sum_{i=0}^{n}z_{i,n}(u)t^{n+2i}\;,
\end{equation}
where the coefficients $z_{i,n}$ depend on the variable $u=1-D/2-\beta$.
For convenience, the polynomials (\ref{41}) are listed, up to the sixth order, in Appendix \ref{app1}.

It is not difficult to see that the evaluation of $A_{-1}(s)$, $A_{0}(s)$ and $A_{i}(s)$
and their first derivative at zero, in the Robin case, follows exactly the same lines as for the Dirichlet case
once the coefficients $x_{i,n}$ are replaced with $z_{i,n}$ \cite{bordag96,bordag96a}.
We will focus our attention on the computation of the function $Z_{R}(s)$ which presents
slight modifications from the Dirichlet case. In what follows the subscript $R$ will denote the Robin case.
We write $Z_{R}(s)$, as we did before, in the form
\begin{equation}
Z_{R}(s)=\sum_{\nu} d(\nu)Z_{\nu,R}(s)\;.
\end{equation}
By taking the derivative of $Z_{\nu,R}(s)$ and setting $s=0$ one obtains
\begin{eqnarray}\label{42}
&&Z_{\nu,R}^{\prime}(0)=\ln\Gamma(\nu+1)+\nu-\nu\ln\nu-\frac{1}{2}\ln 2\pi\nu-\ln\left(1+\frac{u}{\nu}\right)\\
&-&m^{2}\left\{\frac{1}{4(\nu+1)}-\frac{1}{4\nu}-\frac{1}{4\nu^{2}}+\frac{1}{2(\nu+1)(\nu+u)}\right\}
+\sum_{n=1}^{d}\nu^{-n}M_{n}\left(\left(1+\frac{m^{2}}{\nu^{2}}\right)^{-\frac{1}{2}},u\right)+O(m^{4})\;.\nonumber
\end{eqnarray}
By expanding the last term in equation (\ref{42}) in powers of the mass up to the term $m^{2}$ we obtain
\begin{eqnarray}
\sum_{n=1}^{d}\nu^{-n}M_{n}\left(\left(1+\frac{m^{2}}{\nu^{2}}\right)^{-\frac{1}{2}},u\right)&=&\sum_{n=1}^{d}\frac{D_{n}(1)}{\nu^{n}}+\sum_{n=1}^{d}\frac{(-1)^{n+1}}{n}\left(\frac{u}{\nu}\right)^{n}\nonumber\\
&-&m^{2}\sum_{n=1}^{d}\nu^{-n-2}\left[\frac{nD_{n}(1)}{2}+\sum_{b=0}^{n}z_{n,b}b+\frac{(-1)^{n+1}}{2}u^{n}\right]+O(m^{4})\;,
\end{eqnarray}
where we have used the relations \cite{bordag96b,kirsten01}
\begin{eqnarray}
M_{n}(1,0)&=&D_{n}(1)\;,\\
M_{n}(1,u)-M_{n}(1,0)&=&(-1)^{n+1}\frac{u^{n}}{n}\;.
\end{eqnarray}

At this point, by comparing the small $z$ expansion of $\ln[uI_{\nu}(\nu z)+\nu zI^{\prime}_{\nu}(\nu z)]$ with its Olver
expansion we get the following useful relation
\begin{equation}
M^{\prime}_{n}(1,u)=D_{n}^{\prime}(1)+(-1)^{n+1}\sum_{k=0}^{n}u^{k}\;.
\end{equation}
This relation allows us to write the expression for $Z_{\nu,R}^{\prime}(0)$ in (\ref{42}) in the form
\begin{eqnarray}
Z_{\nu,R}^{\prime}(0)&=&Z_{\nu}^{\prime}(0)-\ln\left(1+\frac{u}{\nu}\right)+\sum_{n=1}^{d}\frac{(-1)^{n+1}}{n}\left(\frac{u}{\nu}\right)^{n}\nonumber\\
&-&\frac{m^{2}}{2}\left[\frac{1}{(\nu+1)(\nu+u)}-\frac{1}{\nu^{2}}+\frac{1}{\nu^{2}}\sum_{n=1}^{d}\frac{(-1)^{n+1}}{\nu^{n}}\left(\sum_{k=0}^{n}u^{k}\right)
\right]+O(m^{4})\;.
\end{eqnarray}
One can clearly see from the previous result that in order to study the Robin case we only need to consider the new additional terms.
We would like to point out that this property has already been noticed and utilized in the case of the functional determinant for
massless scalar fields in \cite{bordag96}. Here, we have found that the same feature appears also for the $m^{2}$ correction.
With the last remark in mind, let us define
\begin{equation}\label{43}
N(u)=\sum d(\nu)\left[-\ln\left(1+\frac{u}{n}\right)+\sum_{n=1}^{d}\frac{(-1)^{n+1}}{n}\left(\frac{u}{\nu}\right)^{n}\right]\;,
\end{equation}
and
\begin{equation}
P(u)=\sum d(\nu)\left[\frac{1}{(\nu+1)(\nu+u)}-\frac{1}{\nu^{2}}+\frac{1}{\nu^{2}}\sum_{n=1}^{d}\frac{(-1)^{n+1}}{\nu^{n}}\left(\sum_{k=0}^{n}u^{k}\right)\right]\;.
\end{equation}

The defining expression for $N(u)$ can be evaluated similarly to the procedure utilized in the Dirichlet case.
The calculation for $N(u)$ is shown in detail in \cite{bordag96} and the final result can be explicitly written as
\begin{equation}
N(u)=\zeta^{\prime}_{\mathscr{N}}(0,u)-\frac{1}{2}\zeta_{\mathscr{N}}^{\prime}(0)+\sum_{k=1}^{d}\frac{(-1)^{k+1}}{k}u^{k}
\left[2\textrm{Res}\,\zeta_{\mathscr{N}}\left(\frac{k}{2}\right)\big(\Psi(k)+\gamma\big)+\textrm{FP}\,\zeta_{\mathscr{N}}\left(\frac{k}{2}\right)\right]\;,
\end{equation}
where $\zeta^{\prime}_{\mathscr{N}}(z,u)$ has been defined in (\ref{33b}).

Let us now turn our attention to the term $P(u)$. By utilizing the integral representation (\ref{29})
and by noticing that
\begin{equation}
\sum_{n=0}^{\infty}\frac{(-1)^{n}}{\Gamma(n+2)}\left(\sum_{k=0}^{n}u^{k}\right)t^{n+2}=\frac{t}{u-1}\left(e^{-t}-e^{-ut}\right)\;,
\end{equation}
it is not difficult to prove that
\begin{equation}\label{44}
\frac{1}{(\nu+1)(\nu+u)}=\frac{1}{(u-1)}\int_{0}^{\infty}dt\left(e^{-(\nu+1)t}-e^{-(\nu+u)t}\right)\;.
\end{equation}
At this point an explicit expression for $P(u)$ can be given, namely
\begin{equation}
P(u,z)=\sum d(\nu)\int_{0}^{\infty}dt\,t^{z-1}e^{-\nu t}\left[\frac{te^{-t}}{u-1}-\frac{t e^{-ut}}{u-1}-t^{2}-\sum_{n=1}^{d}\frac{(-1)^{n}}{\Gamma(n+2)}\left(\sum_{k=0}^{n}u^{k}\right)t^{n+2}\right]\;,
\end{equation}
where we have introduced the parameter $z$ in order to regularize the integral.
By using the definition (\ref{33b}) it is not difficult to see that
\begin{eqnarray}
P(u,z)&=&\frac{1}{u-1}\Gamma(z+1)\zeta_{\mathscr{N}}(z+1,1)-\frac{1}{u-1}\Gamma(z+1)\zeta_{\mathscr{N}}(z+1,u)\nonumber\\
&-&\sum_{n=1}^{d}\frac{(-1)^{n}}{\Gamma(n+2)}\left(\sum_{k=0}^{n}u^{k}\right)\Gamma(z+n+2)\zeta_{\mathscr{N}}\left(\frac{z+n+2}{2}\right)-\Gamma(z+2)\zeta_{\mathscr{N}}\left(\frac{z+2}{2}\right)\;.
\end{eqnarray}
As before, we need to take the limit as $z$ approaches zero of the last expression. By recalling the analytic structure
of the spectral $\zeta$-function on the base manifold $\mathscr{N}$, one obtains
\begin{eqnarray}\label{45}
P(u,0)&=&\frac{1}{u-1}\left[\textrm{FP}\,\zeta_{\mathscr{N}}(1,1)-\textrm{FP}\,\zeta_{\mathscr{N}}(1,u)\right]-2(1-\gamma)\textrm{Res}\,\zeta_{\mathscr{N}}(1)-\textrm{FP}\,\zeta_{\mathscr{N}}(1)\nonumber\\
&-&\sum_{n=1}^{d}(-1)^{n}\left(\sum_{k=0}^{n}u^{k}\right)\left[\textrm{FP}\,\zeta_{\mathscr{N}}\left(\frac{n+2}{2}\right)+2\Psi(n+2)\textrm{Res}\,\zeta_{\mathscr{N}}\left(\frac{n+2}{2}\right)\right]\nonumber\\
&-&2\gamma\sum_{n=0}^{d}(-1)^{n}\textrm{Res}\,\zeta_{\mathscr{N}}\left(\frac{n+1}{2}\right)\left(\frac{1-u^{n}}{u-1}\right)+\lim_{z\to 0}\frac{2}{z}\Bigg[\sum_{n=0}^{d}(-1)^{n}\textrm{Res}\,\zeta_{\mathscr{N}}\left(\frac{n+1}{2}\right)\left(\frac{1-u^{n}}{u-1}\right)\nonumber\\
&-&\sum_{n=1}^{d}(-1)^{n}\left(\sum_{k=0}^{n}u^{k}\right)\textrm{Res}\,\zeta_{\mathscr{N}}\left(\frac{n+2}{2}\right)-\textrm{Res}\,\zeta_{\mathscr{N}}\left(1\right)
\Bigg]\;,
\end{eqnarray}
where in order to obtain this formula we have used the fact, which is straightforward to prove, that
\begin{equation}
\textrm{Res}\,\zeta_{\mathscr{N}}(1,u)=2\sum_{k=0}^{d}(-1)^{k}u^{k}\textrm{Res}\,\zeta_{\mathscr{N}}\left(\frac{k+1}{2}\right)\;.
\end{equation}
In the expression that multiplies $1/z$ in (\ref{45}), one can notice that the terms corresponding to $n=0$ vanish identically,
the terms with $n=1$ cancel the residue of $\zeta_{\mathscr{N}}(s)$ at $s=1$, and the remaining terms of the sum with $n\geq 2$
vanish as well due to the following identity
\begin{equation}
\sum_{n=0}^{k}u^{k}=\frac{u^{k+1}-1}{u-1}\;.
\end{equation}
As a consequence, the potentially divergent term in $1/z$ vanishes identically.

We can finally write down the expression for the small mass expansion of the functional determinant of the Laplacian
for Robin boundary conditions as follows
\begin{eqnarray}\label{46}
\zeta_{\mathscr{M},R}^{\prime}(0)&=&\zeta^{\prime}_{\mathscr{N}+1}(0)+\zeta^{\prime}_{\mathscr{N}}(0,u)+\zeta_{\mathscr{N}}\left(-\frac{1}{2}\right)\ln\,2+2\sum_{n=1}^{d}\textrm{Res}\,\zeta_{\mathscr{N}}\left(\frac{n}{2}\right)\left(\sum_{k=0}^{n-1}\frac{1}{k}\right)M_{n}(1)\nonumber\\
&+&2\sum_{n=1}^{d}\textrm{Res}\,\zeta_{\mathscr{N}}\left(\frac{n}{2}\right)\int_{0}^{1}dt\,\frac{M_{n}(t,u)-t^{2}M_{n}(1,u)}{t(1-t^{2})}\nonumber\\
&+&m^{2}\Bigg\{-\frac{1}{4}\textrm{FP}\,\zeta_{\mathscr{N}}(1,1)-\frac{1}{2(u-1)}\big[\textrm{FP}\,\zeta_{\mathscr{N}}(1,1)-\textrm{FP}\,\zeta_{\mathscr{N}}(1,u)\big]+\frac{1}{2}\textrm{Res}\,\zeta_{\mathscr{N}}(1)\nonumber\\
&+&\frac{1}{4}\left(2\ln\,2+1\right)\textrm{Res}\,\zeta_{\mathscr{N}}\left(\frac{1}{2}\right)+\sum_{n=1}^{d}\frac{1}{n}\textrm{Res}\,\zeta_{\mathscr{N}}\left(\frac{n}{2}+1\right)\left[\zeta_{R}(-n)+(-1)^{n+1}u^{n}\right]\nonumber\\
&+&\sum_{n=1}^{d}(-1)^{n}\textrm{Res}\,\zeta_{\mathscr{N}}\left(\frac{n}{2}+1\right)\left(\frac{2u^{n+1}-u-1}{2(u-1)}\right)\left(\frac{2n+1}{n(n+1)}+\sum_{k=1}^{n-1}\frac{1}{k}\right)\nonumber\\
&-&\sum_{n=1}^{d}\textrm{Res}\,\zeta_{\mathscr{N}}\left(\frac{n}{2}+1\right)\int_{0}^{1}dt\,\frac{M_{n}^{\prime}(t,u)-tM_{n}^{\prime}(1,u)}{1-t^{2}}
\Bigg\}+O(m^{4})\;.
\end{eqnarray}
Here once again we would like to point out that the terms of zeroth order in the mass for $\zeta_{\mathscr{M},R}^{\prime}(0)$
coincide with the results obtained in \cite{bordag96}. The terms proportional to $m^{2}$ are, instead, the \emph{new} massive corrections
to the scalar Laplacian on the generalized cone for Robin boundary conditions.

The expressions for $\zeta_{\mathscr{M}}^{\prime}(0)$ for Dirichlet and Robin boundary conditions obtained
in (\ref{38}) and (\ref{46}) contain the spectral $\zeta$-function and its first derivative on the base manifold $\mathscr{N}$.
These expression represent a very general result, holding for any smooth base manifold $\mathscr{N}$ and in dimension $D$.
Without specifying the manifold $\mathscr{N}$ one cannot go further than (\ref{38}) and (\ref{46}) in the evaluation of the functional determinant.
In the next section we will specify the base manifold to be a $d$-dimensional ball. This case is of particular importance
because the spectral $\zeta$-function on the $d$-dimensional ball can be evaluated explicitly in terms of the
Barnes $\zeta$-function \cite{bordag96,chang93}.

\section{$d$-Dimensional sphere as Base Manifold $\mathscr{N}$}

In this section we will assume that the base manifold is a $d$-dimensional sphere. In this case the
relevant Bessel function index is
\begin{equation}
\nu=\left(l+\frac{d-1}{2}\right)\;,
\end{equation}
and the eigenfunctions are hyperspherical harmonics with the degeneracy
\begin{equation}
d(l)=(2l+d-1)\frac{(l+d-2)!}{l!(d-1)!}\;.
\end{equation}
By using the definition (\ref{10}), we can write $\zeta_{\mathscr{N}}(s)$ as
\begin{equation}\label{47}
\zeta_{\mathscr{N}}(s)=\sum_{l=0}^{\infty}(2l+d-1)\frac{(l+d-2)!}{l!(d-1)!}\left(l+\frac{d-1}{2}\right)^{-2s}\;.
\end{equation}
It is not difficult to show, with some algebraic manipulations on the factorials, that the $\zeta$-function
in (\ref{47}) can be written as a sum of Barnes $\zeta$-functions \cite{bordag96,chang93}
\begin{equation}\label{47a}
\zeta_{\mathscr{N}} (s) =\zeta_{\mathscr{B}}\left(2s,\frac{d+1}{2}\right)+\zeta_{\mathscr{B}}\left(2s,\frac{d-1}{2}\right)\;,
\end{equation}
where the Barnes $\zeta$-function is defined as \cite{barnes03,dowker05}
\begin{equation}
\zeta_{\mathscr{B}}(s,a|\vec{r})=\sum_{\vec{m}=0}^{\infty}\frac{1}{(a+\vec{m}\cdot\vec{r})^{s}}\;,
\end{equation}
valid for ${\rm Re}(s)>d$ where $\vec{m}$ and $\vec{r}$ are $d$-dimensional vectors, and where the notation $\zeta_{\mathscr{B}} (s,a|\vec 1)=\zeta_{\mathscr{B}} (s,a)$ has been used.

In our case, we consider the integral representation of the Barnes $\zeta$-function, namely \cite{bordag96}
\begin{equation}\label{48}
\zeta_{\mathscr{B}}(s,a)=\frac{i\Gamma(1-s)}{2\pi}\int_{L}dz\,\frac{e^{z\left(\frac{d}{2}-a\right)}(-z)^{s-1}}{2^{d}\sinh^{d}\left(\frac{z}{2}\right)}\;,
\end{equation}
where $L$ represents the Hankel contour. By using this integral representation in the expression (\ref{47a})
we can write $\zeta_{\mathscr{N}}(s) $ as
\begin{equation}\label{49a}
\zeta_{\mathscr{N}}(s)=\frac{i\Gamma(1-2s)}{2\pi}\int_{L}dz\,\frac{(-z)^{2s-1}\cosh\left(\frac{z}{2}\right)}{2^{d-1}\sinh^{d}\left(\frac{z}{2}\right)}\;.
\end{equation}
With the help of a simple change of variables, $z/2\to z$, we can rewrite the result in (\ref{49a}) to obtain \cite{bordag96}
\begin{equation}\label{50}
\zeta_{\mathscr{N}}(s)=(-1)^{2s-2}\frac{i\Gamma(2-2s)}{2\pi(d-1)}2^{2s+1-d}\sum_{\nu=0}^{\infty}\frac{D_{\nu}^{(d-1)}}{\nu!}\int_{L}dz\,z^{2s-d-1+\nu}\;,
\end{equation}
where the coefficients $D_{\nu}^{(d-1)}$ are defined as \cite{bordag96,chodos84}
\begin{equation}
\left(\frac{z}{\sinh z}\right)^{d-1}=\sum_{\nu=0}^{\infty}D_{\nu}^{(d-1)}\frac{z^{\nu}}{\nu!}\;.
\end{equation}

In particular, we will need to compute, for Dirichlet boundary conditions, the residues of $\zeta_{\mathscr{N}}(s)$ at the points $s=m/2$ with $m$ being a positive integer.
It is easy to see that the integral in (\ref{50}) vanishes identically unless $\nu=d-m$ where the integrand
has simple poles. By using the residue method we obtain \cite{bordag96}
\begin{equation}\label{49}
\textrm{Res}\,\zeta_{\mathscr{N}}\left(\frac{m}{2}\right)=\frac{2^{m-d}D_{d-m}^{(d-1)}}{(d-1)(m-2)!(d-m)!}\;,
\end{equation}
valid for $m\geq 2$ and $d\geq m$. Another particular value that we will need in the subsequent calculations
is $\zeta_{\mathscr{N}}(-1/2)$. Again from (\ref{50}) it is straightforward to show that
\begin{equation}\label{51}
\zeta_{\mathscr{N}}\left(-\frac{1}{2}\right)=\frac{2^{1-d}D_{d+1}^{(d-1)}}{(d-1)(d+1)!}\;.
\end{equation}
In addition to the result (\ref{49}), we will need, in the case of Robin boundary conditions, the residues of $\zeta_{\mathscr{N}}(s)$
at $s=m/2+1$.
These follow immediately from (\ref{49}) by replacing $m$ with $m+2$, the result then being valid for $m\geq 0$ and $d\geq m+2$.

For the $d$-dimensional ball we also have that
\begin{equation}
\zeta_{\mathscr{N}+1}(s)=\sum_{l=0}^{\infty}e(l)\left(l+\frac{d+1}{2}\right)^{-s}\;,
\end{equation}
where
\begin{equation}
e(l)=(2l+d)\frac{(l+d-1)!}{l!d!}.\nonumber
\end{equation}
By writing \cite{bordag96}
\begin{equation}\label{52}
e(l)= \sum_{\alpha=0}^{d}e_{\alpha}\left(l+\frac{d+1}{2}\right)^{\alpha}\;,
\end{equation}
which defines $e_\alpha$, we have that
\begin{equation}
\zeta_{\mathscr{N}+1}(s)=\sum_{\alpha=0}^{d}e_{\alpha}\zeta_{H}\left(s-\alpha,\frac{d+1}{2}\right)\;,
\end{equation}
where the coefficients $e_{\alpha}$ depend on the dimension $d$ and are determined from the equation (\ref{52}), and $\zeta_{H}$ represents the
Hurwitz $\zeta$-function.
In particular, we have
\begin{equation}
\zeta^{\prime}_{\mathscr{N}+1}(0)=\sum_{\alpha=0}^{d}e_{\alpha}\zeta^{\prime}_{H}\left(-\alpha,\frac{d+1}{2}\right)\;.
\end{equation}

For our calculations we will also need the following expression
\begin{equation}\label{52a}
\zeta_{\mathscr{N}}(s,u)=\sum_{\alpha=0}^{d-1}e_{\alpha}(u)\zeta_{H}\left(s-\alpha,\frac{d-1}{2}+u\right)\;,
\end{equation}
where the coefficients $e_{\alpha}(u)$ are easily found by utilizing the equality \cite{bordag96}
\begin{equation}
d(l)=(2l+d-1)\frac{(l+d-2)!}{l!(d-1)!}=\sum_{\alpha=0}^{d-1}e_{\alpha}(u)\left(l+\frac{d-1}{2}+u\right)^{\alpha}\;.
\end{equation}

\section{Specific Dimensions for Dirichlet Boundary Conditions}

By utilizing the result for $\zeta^{\prime}_{\mathscr{M}}(0)$ and the particular form obtained for $\zeta_{\mathscr{N}}(s)$ and
$\zeta_{\mathscr{N}+1}(s)$ when the base manifold is a sphere, we can get explicit expressions for specific dimensions.
Here, we will present the results for $d=2,3,4,5$. Obviously, results for higher dimensions can be extracted with some amount of work from
the formulas presented in the previous sections.

For the base manifold of dimension $d=2$ we obtain
\begin{eqnarray}
\zeta_{\mathscr{M}}^{\prime}(0)=-\frac{3}{32}-\frac{1}{12}\ln\,2+\frac{1}{3}\zeta_{R}^{\prime}(-1)-\frac{3}{4}\zeta_{R}^{\prime}(-2)+m^{2}\left[-1+\frac{1}{2}(\gamma+2\ln\,2)\right]+O(m^{4})\;.
\end{eqnarray}

For the base manifold of dimension $d=3$ we obtain
\begin{eqnarray}
\zeta_{\mathscr{M}}^{\prime}(0)&=&\frac{173}{30240}+\frac{1}{90}\ln\,2+\frac{1}{6}\zeta_{R}^{\prime}(-1)-\frac{1}{2}\zeta_{R}^{\prime}(-2)+\frac{1}{3}\zeta_{R}^{\prime}(-3)\nonumber\\
&+&m^{2}\left[-\frac{5}{24}+\frac{1}{4}(\ln\,2-\gamma)\right]+O(m^{4})\;.
\end{eqnarray}

For the base manifold of dimension $d=4$ we obtain
\begin{eqnarray}
\zeta_{\mathscr{M}}^{\prime}(0)&=&\frac{47}{9216}+\frac{17}{2880}\ln\,2-\frac{1}{48}\zeta_{R}^{\prime}(-1)-\frac{1}{32}\zeta_{R}^{\prime}(-2)+\frac{7}{48}\zeta_{R}^{\prime}(-3)-\frac{5}{64}\zeta_{R}^{\prime}(-4)\nonumber\\
&+&m^{2}\left[-\frac{7}{576}+\frac{1}{16}(\gamma+2\ln\,2)\right]+O(m^{4})\;,
\end{eqnarray}

For the base manifold of dimension $d=5$ we obtain
\begin{eqnarray}
\zeta_{\mathscr{M}}^{\prime}(0)&=&-\frac{4027}{6486480}-\frac{1}{576}\ln\,2-\frac{1}{60}\zeta_{R}^{\prime}(-1)+\frac{1}{24}\zeta_{R}^{\prime}(-2)-\frac{1}{24}\zeta_{R}^{\prime}(-4)\nonumber\\
&+&\frac{1}{60}\zeta_{R}^{\prime}(-5)-m^{2}\frac{143}{6048}+O(m^{4})\;.
\end{eqnarray}
In order to obtain the above formulas for even dimensions we have exploited the formula
\begin{eqnarray}\label{53}
\zeta_{H}^{\prime}\left(s,q+\frac{1}{2}\right)=2^{s}\ln\,2\left[\zeta_{R}(s)-\sum_{n=1}^{2q-1}\frac{1}{n^{s}}\right]+2^{s}\left[\zeta^{\prime}_{R}(s)+\sum_{n=1}^{2q-1}\frac{\ln\,n}{n^{s}}\right]-\zeta^{\prime}_{R}(s)-\sum_{n=1}^{q-1}\frac{\ln\,n}{n^{s}}\;,
\end{eqnarray}
valid for any integer $q\geq 0$, while for odd dimensions we have used
\begin{equation}\label{54}
\zeta_{H}^{\prime}(s,z)=\zeta_{R}^{\prime}(s)+\sum_{n=0}^{z-2}\frac{\ln(n+1)}{(n+1)^{s}}\;,
\end{equation}
which holds for any integer $z\geq 2$.
We would like to point out that the results for the zeroth order in mass obtained
for $d=2,3,4,5$ coincide with the ones obtained in \cite{bordag96,kirsten01}.

\section{Specific Dimensions for Robin Boundary Conditions}

For Robin boundary conditions we consider the general result obtained in (\ref{46}) and specialize it
to the case in which the base manifold is a sphere of dimension $d=2,3,4,5$. In addition to the
spectral functions $\zeta_{\mathscr{N}}(s)$ and $\zeta_{\mathscr{N}+1}(s)$ obtained in the previous sections,
we will also need $\textrm{FP}\zeta_{\mathscr{N}}(1,u)$. It is straightforward to show,
by using (\ref{52a}), that
\begin{equation}
\textrm{FP}\zeta_{\mathscr{N}}(1,u)=-e_{0}(u)\Psi\left(\frac{d-1}{2}+u\right)-\sum_{\alpha=0}^{d-2}e_{\alpha+1}(u)\frac{B_{\alpha+1}\left(\frac{d-1}{2}+u\right)}{\alpha+1}\;,
\end{equation}
where $B_{\alpha}(z)$ are the Bernoulli polynomials.

For the base manifold of dimension $d=2$ we obtain
\begin{eqnarray}
\zeta^{\prime}_{\mathscr{M},R}(0)&=&\frac{1}{32}-\frac{1}{6}\ln\,2+\frac{u}{2}-\frac{1}{2}
\zeta^{\prime}_{R}(-1)-\frac{3}{4}\zeta^{\prime}_{R}(-2)
-2u\ln\Gamma\left(\frac{1}{2}+u\right)+2\int_{0}^{u}dx\,\ln\Gamma\left(\frac{1}{2}+x\right)\nonumber\\
&+&m^{2}\left[\frac{1}{2}(\gamma+2\ln\,2)\frac{u+1}{u-1}-\frac{u+1}{u-1}+\frac{u}{u-1}\Psi\left(\frac{1}{2}+u\right)\right]+O(m^{4})\;.
\end{eqnarray}

For the base manifold of dimension $d=3$ we have
\begin{eqnarray}
\zeta^{\prime}_{\mathscr{M},R}(0)&=&\frac{11}{4320}+\frac{1}{90}\ln\,2+\frac{u}{30}-\frac{5}{12}u^{2}-\frac{u^{3}}{3}+\frac{1}{6}\zeta_{R}^{\prime}(-1)+\frac{1}{2}\zeta_{R}^{\prime}(-2)+\frac{1}{3}\zeta_{R}^{\prime}(-3)\nonumber\\
&+&u^{2}\ln\Gamma(1+u)-2\int_{0}^{u}dx\,x\ln\Gamma(1+x)+m^{2}\Bigg\{-\frac{5}{24}-\frac{\gamma}{4}-\frac{1}{4}\ln\,2-\frac{u}{2}\ln\,2\nonumber\\
&-&\frac{u}{4}-\frac{1}{2(u-1)}\left[1+\gamma-\frac{u}{2}+u^{2}\left(\Psi(u+1)-\frac{3}{2}\right)\right]\Bigg\}+O(m^{4})\;.
\end{eqnarray}

For the base manifold of dimension $d=4$ we have
\begin{eqnarray}
\zeta^{\prime}_{\mathscr{M},R}(0)&=&-\frac{61}{46080}-\frac{11}{576}u-\frac{1}{16}u^{2}+\frac{11}{72}u^{3}+\frac{1}{24}u^{4}+\frac{7}{720}\ln\,2+\frac{1}{48}\zeta_{R}^{\prime}(-1)\nonumber\\
&-&\frac{1}{32}\zeta_{R}^{\prime}(-2)-\frac{7}{48}\zeta_{R}^{\prime}(-3)-\frac{5}{64}\zeta_{R}^{\prime}(-4)+\frac{u}{12}\ln\Gamma\left(\frac{3}{2}+u\right)-\frac{1}{3}u^{3}\ln\Gamma\left(\frac{3}{2}+u\right)\nonumber\\
&-&\frac{1}{12}\int_{0}^{u}dx\,\ln\Gamma\left(\frac{3}{2}+x\right)+\int_{0}^{u}dx\,x^{2}\ln\Gamma\left(\frac{3}{2}+x\right)+m^{2}\Bigg\{\frac{13}{576}+\frac{5}{36}u+\frac{2}{9}u^{2}\nonumber\\
&+&\frac{\gamma}{16}+\frac{1}{8}\ln\,2-\frac{1}{2(u-1)}\Bigg[-\frac{1}{24}-\frac{\gamma}{4}-\frac{1}{2}\ln\,2-\frac{17}{72}u+\frac{u^{2}}{3}+\frac{11}{18}u^{3}\nonumber\\
&-&\frac{u}{12}(4u^{2}-1)\Psi\left(\frac{3}{2}+u\right)\Bigg]\Bigg\}+O(m^{4})\;.\nonumber\\
\end{eqnarray}

And lastly, for the base manifold of dimension $d=5$ we have
\begin{eqnarray}
\zeta^{\prime}_{\mathscr{M},R}(0)&=&-\frac{9479}{32432400}-\frac{u}{315}+\frac{517}{15120}u^{2}+\frac{83}{1512}u^{3}-\frac{19}{480}u^{4}-\frac{u^{5}}{45}-\frac{1}{756}\ln\,2-\frac{u^{3}}{36}\ln\,2\nonumber\\
&+&\frac{u^{5}}{60}\ln\,2-\frac{1}{60}\zeta_{R}^{\prime}(-1)-\frac{1}{24}\zeta_{R}^{\prime}(-2)+\frac{1}{24}\zeta_{R}^{\prime}(-4)+\frac{1}{60}\zeta_{R}^{\prime}(-5)-\frac{1}{12}u^{2}\ln\Gamma(u+2)\nonumber\\
&+&\frac{1}{12}u^{4}\ln\Gamma(u+2)+\frac{1}{6}\int_{0}^{u}dx\,x\ln\Gamma(x+2)-\frac{1}{3}\int_{0}^{u}dx\,x^{3}\ln\Gamma(x+2)\nonumber\\
&+&m^{2}\Bigg\{-\frac{13}{4320}+\frac{u}{40}-\frac{u^{2}}{96}-\frac{u^{3}}{32}-\frac{u^{2}}{24}\ln\,2-\frac{u^{3}}{24}\ln\,2-\frac{1}{2(u-1)}\Bigg[\frac{1}{24}+\frac{u}{24}\nonumber\\
&+&\frac{31}{144}u^{2}-\frac{u^{3}}{8}-\frac{25}{144}u^{4}+\frac{u^{2}}{12}(u^{2}-1)\Psi(u+2)
\Bigg]
\Bigg\}+O(m^{4})\;.
\end{eqnarray}
We would like to mention, once again, that the results of zeroth order in mass obtained in this section
coincide with the ones obtained in \cite{bordag96,kirsten01}. Moreover, it is worth noticing that although the above results
for $\zeta^{\prime}_{\mathscr{M},R}(0)$ are finite in the limit as $u\to1$, corresponding to Neuman boundary conditions, they do not reproduce the correct result.
This is due to the fact that, when studying Neuman boundary conditions, particular care is needed with the zero modes which have to be dealt with separately \cite{bordag96b}.
The terms involving integrals over $\ln \Gamma (w)$ could be given more explicitly in terms of the zeta function of Riemann and its derivative, \cite{cag}, but the chosen form is more compact.

\section{Conclusions}
The article continues the analysis of the functional determinant of the Laplacian on the generalized cone started in \cite{bordag96,bordag96b,stuart}. Whereas in these references results were given for $m=0$ only, this article provides a way to evaluate a systematic small-$m$ expansion in powers of $m^2$. The leading order correction for Dirichlet and Robin boundary conditions on the generalized cone are given in (\ref{38}) and (\ref{46}). Specializing to the example of a ball the results in Sections 6 and 7 are found. Higher orders can be computed as needed. Combined with large mass expansions coming from the heat kernel it is hoped that simple interpolating techniques are sufficient to find results for massive functional determinants essentially for all masses.

{\bf Acknowledgements:} KK is supported by the National Science Foundation Grant
PHY-0757791.

\appendix

\section{Polynomials $D_{n}(t)$ and $M_{n}(t)$ up to the order $n=6$}\label{app1}
We list, here, the polynomials $D_{n}(t)$ up to the sixth order. By utilizing the expression (\ref{21})
and the recurrence relation (\ref{16}) it is straightforward to obtain
\begin{eqnarray}
D_{1}(t)&=&\frac{1}{8}t-\frac{5}{24}t^{3}\;,\\
D_{2}(t)&=&\frac{1}{16}t^2-\frac{3 }{8}t^4+\frac{5 }{16}t^6\;,\\
D_{3}(t)&=&\frac{25 }{384}t^3-\frac{531 }{640}t^5+\frac{221 }{128}t^7-\frac{1105 }{1152}t^9\;,\\
D_{4}(t)&=&\frac{13}{128}t^4-\frac{71 }{32}t^6+\frac{531 }{64}t^8-\frac{339}{32}t^{10}+\frac{565 }{128}t^{12}\;,\\
D_{5}(t)&=&\frac{1073}{5120}t^5-\frac{50049 }{7168}t^7+\frac{186821 }{4608}t^9-\frac{44899 }{512}t^{11}+\frac{82825 }{1024}t^{13}-\frac{82825 }{3072}t^{15}\;,\\
D_{6}(t)&=&\frac{103}{192}t^6-\frac{405 }{16}t^8+\frac{1677 }{8}t^{10}-\frac{5389 }{8}t^{12}+\frac{65385 }{64}t^{14}-\frac{11805 }{16}t^{16}+\frac{19675 }{96}t^{18}\;.
\end{eqnarray}

The polynomials $M_{n}(t,u)$ follow by using the expression (\ref{40}) together with the recurrence relation (\ref{39}).
One can easily find
\begin{eqnarray}
M_{1}(t,u)&=&-\frac{3}{8} t+\frac{7}{24} t^3+t u\;,\\
M_{2}(t,u)&=&-\frac{3}{16}t^2+\frac{5}{8}t^4-\frac{7}{16}t^6+\frac{u}{2}t^2-\frac{u}{2}t^4-\frac{u^2}{2}t^2\;,\\
M_{3}(t,u)&=&-\frac{21}{128} t^3+\frac{869}{640} t^5-\frac{315}{128} t^7+\frac{1463}{1152} t^9+\frac{3u}{8} t^3 -\frac{5u}{4} t^5 +\frac{7u}{8} t^7 -\frac{u^2}{2}t^3\nonumber\\
 &+&\frac{u^2}{2}t^5 +\frac{u^3}{3}t^3 \;, 
\end{eqnarray}
\begin{eqnarray}
M_{4}(t,u)&=&-\frac{27}{128} t^4+\frac{109}{32} t^6-\frac{733}{64} t^8+\frac{441}{32} t^{10}-\frac{707}{128} t^{12}
+\frac{3u}{8} t^4 -\frac{23u}{8} t^6 +\frac{41u}{8} t^8\nonumber\\
&-&\frac{21u}{8} t^{10} -\frac{u^2}{2}t^4
+\frac{3u^2}{2} t^6 -t^8 u^2+\frac{u^3}{2}t^4 -\frac{u^3}{2}t^6 -\frac{u^4}{4}t^4 \;,\\
M_{5}(t,u)&=&-\frac{1899}{5120} t^5+\frac{72003}{7168} t^7-\frac{247735}{4608} t^9+\frac{56761}{512} t^{11}-\frac{101395}{1024} t^{13}
+\frac{495271}{15360} t^{15}+\frac{63u}{128} t^5\nonumber\\
&-&\frac{233u}{32} t^7 +\frac{1537u}{64} t^9 -\frac{917u}{32} t^{11} +\frac{1463u}{128} t^{13}
-\frac{9u^2}{16} t^5 +\frac{59u^2}{16} t^7 -\frac{99u^2}{16} t^9 +\frac{49u^2}{16} t^{11}\nonumber\\
&+&\frac{5u^3}{8} t^5 -\frac{7u^3}{4} t^7
+\frac{9u^3}{8} t^9 -\frac{u^4}{2}t^5 +\frac{u^4}{2}t^7 +\frac{u^5}{5}t^5\;,\nonumber\\
M_{6}(t,u)&=&-\frac{27}{32} t^6+\frac{69}{2} t^8-\frac{17163}{64} t^{10}+\frac{4973}{6} t^{12}-\frac{9789}{8} t^{14}
+\frac{3465}{4} t^{16}-\frac{45493}{192} t^{18}+\frac{27u}{32} t^6 \nonumber\\
&-&\frac{681u}{32} t^8 +\frac{1793u}{16} t^{10}
-\frac{3671u}{16} t^{12} +\frac{6531u}{32} t^{14} -\frac{2121u}{32} t^{16} -\frac{3u^2}{4} t^6 +\frac{75u^2}{8} t^8 \nonumber\\
&-&\frac{233u^2}{8} t^{10} +\frac{269u^2}{8} t^{12} -\frac{105u^2}{8} t^{14} +\frac{19u^3}{24} t^6 -\frac{37u^3}{8} t^8
+\frac{59u^3}{8} t^{10} -\frac{85u^3}{24} t^{12} \nonumber\\
&-&\frac{3u^4}{4} t^6 +2 t^8 u^4-\frac{5u^4}{4} t^{10} +\frac{u^5}{2}t^6
-\frac{u^5}{2}t^8 -\frac{u^6}{6}t^6 \;.
\end{eqnarray}

\end{document}